\documentclass[10pt]{article}
\usepackage{amsmath}
\usepackage{graphicx}
\begin{document}
\title{The Practice of Naturalness:\\ A Historical-Philosophical Perspective\thanks{Forthcoming in {\it Foundations of Physics}}}

\author{{Arianna Borrelli\thanks{Institute for Advanced Study on Media Cultures of Computer Simulation (MECS), Leuphana University L\"uneburg, Universitätsallee 1, 21335 L\"uneburg, Germany. E-mail: \mbox{aborrelli@weatherglass.de}}} and  {Elena Castellani\thanks{Department of Humanities and Philosophy, University of Florence, via Bolognese 52, 50139,
Firenze, Italy. E-mail: elena.castellani@unifi.it}}}
\date{}
\maketitle

\section{Introduction: Naturalness from problem to principle}

Ever since mathematization started playing a role in the natural sciences, many have  justified their research choices with the help of more or less sketchy arguments involving preferences for certain types of numbers (e.g. integers, small fractions, primes...), geometric structures or mathematical forms. Such arguments have often been expressed with words like ``simple", ``beautiful", ``plausible/implausible" or ``natural". For the most part arguments of this kind were brought forward in reference to specific contexts and were presented as tentative reflections aimed at pursuing a well-defined goal, rather than as expression of general principles of natural order or scientific methodology. 

Indeed, trying to connect statements about what was seen as natural, simple, or beautiful in different historical-epistemological contexts is not so straightforward and may lead to vague, generic characterizations, as seen in recent attempts to define the alleged successes or failures of the naturalness principle (Giudice 2008, Wells 2015).
Recent discussions in high energy physics and cosmology frame the issue of naturalness as that of a principle reflecting some fundamental feature of natural laws. At the same time, it is assumed that naturalness has guided the developments of high energy physics (and to some extent of cosmology) since the 1980s. Based on these premises, various questions have been discussed by physicists as well as philosophers, such as:

\begin{itemize}

\item[$\bullet$] How can the principle of naturalness be exactly and coherently formulated?

\item[$\bullet$] Was the principle (in one or the other formulation) indeed successful in guiding research?

\item[$\bullet$] Can it be assumed that the naturalness principle (however defined) will productively guide future research, or should it be rejected altogether?

\end{itemize}

\noindent We will not specifically focus on any of these issues, but rather question the premises of their framework. In particular, we will argue that, in high energy physics from the 1970s onward, arguments today subsumed under the ``naturalness principle'' were employed as situated reflections aimed at pursuing specific goals, and not as expressions of some overarching fundamental principle. 

Looking back at the historical record shows how naturalness could be extremely popular on the High-Energy-Physics scene, while at the same time lacking a coherent definition. In fact, a broad range of different, sometimes even incompatible ideas of naturalness peacefully thrived alongside each other, allowing to identify various naturalness problems and their possible solutions (Borrelli 2015). There was no overarching principle guiding the process, but rather a dynamic interplay of problems and solutions. And until less than ten years ago, it was implicitly assumed that ultimately empirical results from experiments (especially the LHC) would eventually help differentiate within the model landscape. It is significant that one of the earliest overviews on the notion of naturalness (Giudice 2008) appeared in a volume devoted to \textit{Perspectives on LHC Physics} published shortly before the start of the LHC machine at CERN.

In fact, until the early 21st century, hardly anyone had seen the vagueness of naturalness as problematic. It was only when the results from the LHC suggested that there might be no problem with the Standard Model after all, that critical attention focused on naturalness, and the {\it a posteriori} construction of a naturalness principle guiding research (astray, for the critics) emerged. 

In what follows we will address the issue of naturalness by focusing on a few case studies. The aim is to show how a number of parallel, at times competing processes of model building combined and conflicted with each other with the result of giving rise to a ``naturalness problem".

\section{``Natural" models in high-energy physics: an overview}

As a first heuristic attempt to grasp the broader historical-epistemological developments in which the issue of naturalness emerged, we will distinguish four historical phases providing a rough orientation for the  analysis of the following Sections.

\textit{First phase (1970-1979):} In this period the Standard Model slowly emerged, the methods of the renormalization group and of spontaneous symmetry breaking were studied, and at times combined to try and predict some values of masses and coupling constants. Although attempts of this kind were often described as simple or elegant, the term ``natural" was increasingly used in this context. Many of these arguments have been a posteriori reconstructed as more or less successful applications of an alleged naturalness principle, but a closer look at the historical sources shows that this was not the case. To support this thesis we offer a critical discussion of Kenneth Wilson's 1971 paper which is often indicated as the locus classicus of the emergence of naturalness (Sections 3 and 4). We will argue that in Wilson's paper neither a naturalness principle nor a naturalness problem appeared.

\textit{Second phase (1980-1985):} By 1980 the Standard Model was fully established, but the search for alternative theories of particles went on, with a broad range of proposals to embed the Standard Model into a more unified theory (e.g. grand unified theories, supersymmetry, technicolor). The main strategy to support one or the other of these proposals was to refer to the existence of problems with the Standard Model, often in connection with the Higgs boson. Models were supported by pointing out distinct, though often vaguely related problems, and the term ``naturalness problem" established itself to denote this broad class. At the time it was clear to most that the problems were often arbitrarily or even incoherently formulated. For example, as we shall see more in detail in section 5, quadratic divergences in the renormalization of the Higgs were regarded as indications that the Standard model was unnatural, although it was known that those divergences disappeared when employing dimensional regularization. Yet this ambiguity was not felt as a real difficulty, since these problems were regarded as situated arguments pursuing a specific goal.

This period was characterized by the publication of three seminal papers on naturalness respectively by Leonard Susskind (1979), Gerhard 't Hooft (1980) and Martin Veltmann (1981). Since these texts have been often discussed, we will only briefly summarize their main claims (Section 5), and rather dwell on a much less known article published in 1985 by physicist Philip Nelson in the popular science journal \textit{American Scientist} (Section 6). Indeed, this paper is apparently the first attempt to offer a systematic discussion of different naturalness criteria and problems. Morever, it marks the moment when naturalness became popular.

\textit{Third phase (1985-2012):} In the 1980s supersymmetry established itself as a most promising solution to the naturalness problem, and LEP results were widely expected to provide evidence in its favor. After LEP results failed to fulfill the expectations, some versions of the naturalness problem were redefined in terms of fine-tuning and continued to serve as a heuristic guideline for model-building, especially, but not only, for supersymmetry. Around 2000, versions of the naturalness problem were introduced which could be solved by newly proposed extra-dimensional theories. Here, we will limit our analysis to a brief sketch of the way in which the naturalness problem, after being connected to supersymmetry during the 1980s, proved to be an effective, flexible tool to guide and motivate research in that field of model building (Section 7).

\textit{Fourth phase (2012 onwards):} This is the period we are in, which Gian Francesco Giudice proposed to characterize as a ``post-naturalness era"  (Giudice 2017). LHC results, confirming the predictions of the Standard Model and providing no evidence of new physics, have started increasingly heated discussions on naturalness.

\section{Kenneth Wilson (1971): the origin of naturalness?}

In 1971 Kenneth Wilson published a paper on ``Renormalization group and strong interactions" which is usually considered as marking the beginning of the naturalness narrative in theoretical physics.\footnote{Recent examples of this claim can be found in Maiani/Bonolis 2017, 648; Arkani-Hamed et al. 2016, 28; Patrignani and Particle Data Group 2016, 173 and 215.} Even Wilson himself, many decades later, referred to this work as originating the idea that massive scalar particles are unnatural, albeit only to remark that his earlier claim "made no sense" (Wilson 2005, 12).  

Apparently, the connection between Wilson's paper and the naturalness issue was established for the first time by  Susskind in his seminal 1979 naturalness paper, where he acknowledged Wilson for ``explaining the reason why scalar fields require unnatural adjustments of bare constants" (Susskind 1979, 2624). Let us note that, while Susskind offered no specific reference to published material, already in Veltmann (1981) we can find an explicit reference to the following lines from Wilson's 1971 paper:  

\begin{quote}
It is interesting to note that there are no weakly coupled scalar particles in nature; scalar particles are the only kind of free particles whose mass does not break either an internal or a gauge symmetry. This discussion can be summarized by saying that mass or symmetry breaking terms must be 'protected' from large corrections at large momenta due to various interactions (electromagnetic, weak or strong). A symmetry-breaking term [...] is protected if, in the renormalization-group equation the right-hand side is proportional to [the coupling constant of the term] (Wilson 1971, 1840).
\end{quote}

\noindent  What did Wilson mean with these words? Was he really stating a naturalness problem or naturalness principle? What were his goals and premises at the time? We shall address these questions by taking a closer, historically contextualized look at Wilson's paper, reconstructing at least in part its goals, arguments and results. As said, our aim is not so much to discover anything new about naturalness, but rather to put these developments in the correct perspective, thus avoiding the current risk of projecting back today's concepts and arguments into past writings.

Before turning to Wilson's paper, it will be useful to situate it historically by recalling the state of particle physics in the early 1970s and the developments which still lay ahead at the time.\footnote{For a detailed treatment of high energy physics in the 1960s and '70s see (Hoddeson et al. 1997; Pickering 1984b).} As is well known, the 1970s represented a turning point for the history of high energy physics, with the emergence and establishment both of the Standard Model and of the concepts and techniques of the renormalization group. Although Abdus Salam and Steven Weinberg had published already in late 1960s the papers which are today regarded as the origin of the Weinberg-Salam model of electroweak interactions, those papers had been largely ignored by the community. The situation changed only in 1971-72 when, thanks to the work of 't Hooft and Veltmann, it became generally accepted that a quantum field theory with a spontaneously broken gauge symmetry could be renormalizable. Interpreting Wilson's words as referring to the Higgs mechanism, as done by (Grinbaum 2012, 620), seems therefore somewhat anachronistic.

The work of 't Hooft and Veltman spurred a focused experimental search for neutral weak currents, evidence for which had often been explained away in the previous years.\footnote{For a detailed discussion of the discovery and non-discovery of weak neutral currents see (Pickering 1984a).}  When those effects were detected at the Gargamelle bubble chamber at CERN in 1973, however, this result was not seen primarily as a specific confirmation of the Weinberg-Salam model, but rather of the validity of its basic ingredients: quantum field theory, non-Abelian gauge symmetry and spontaneous symmetry breaking. Once those principles of model-building were established, the 1970s saw the creation (and disappearance) of a large number of models of electroweak and strong interactions. See for example how Steven Weinberg described the situation when attempting to take stock of all possible theories of broken gauge symmetry:

\begin{quote}

It was hoped that when a finite theory of weak and electromagnetic interactions was finally discovered, it would be unique or at least very constrained. Unfortunately, once one finite theory was written down, it was obvious how to write down many classes of Lagrangians for weak and electromagnetic interactions which were finite. Therefore let us first address ourselves to the problem of determining the possible degrees of freedom in choosing a gauge model (Weinberg 1974, 36).

\end{quote}

\noindent  Eventually, it became accepted that the Weinberg-Salam model in its original form indeed provided the best match for experimental results, while QCD established itself as the theory of strong interactions. 

In those years, people also became increasingly aware of the formal intricacies and possible physical implications of renormalization procedures, especially in cases of spontaneous symmetry breaking, and started exploring them further, often with an aim of deriving from them relationships between - or even predictions for - the observed values of coupling constants and masses.  These explorations usually had a semi-qualitative character and were guided in part by mathematical procedures and in part by analogies with already known cases or with toy models. In this context, the term ``natural" started being employed to characterize models in which the values of observable masses and coupling constants were determined by renormalization conditions or by (broken or unbroken) symmetries. For example,  when discussing the effects of spontaneous symmetry breaking, Weinberg suggested that it would provide a ``natural explanation of the approximate symmetries of nature" (Weinberg 1972, 1698), while Howard Georgi and Sheldon Glashow regarded it as a possible ``natural mechanism for mass hierarchy" (Georgi and Glashow 1972, 2979). Georgi and Abraham Pais even introduced a quite refined notion of ``naturalness" based on the idea that a relation between two parameters in quantum field theory would be ``natural" if, due to effects of spontaneous symmetry breaking, it only received finite radiative corrections (Georgi and Pais 1974, 539-540).\footnote{Georgi and Pais's definition is the earliest systematic use we could find of the term naturalness in particle physics, although its characterization has little in common with later notions.} 

An increasingly important ingredient in this complex landscape was the renormalization group, playing a key role both in the establishment of QCD and in the development of grand unified theories. Today, it provides the backbone for more technical versions of the naturalness principle (Wells 2009). Here Wilson's contribution comes into play: although the key ideas of the renormalization group had been introduced already in the 1950s, he was the first author to systematically expand and employ this approach to derive physical consequences. Indeed, the first in the series of Wilson's seminal papers on this topic was the same one which today is seen as the origin of naturalness as intended today.

\section{Wilson's 1971 paper: goals and results}

In his 1971 paper Wilson, building upon the work by Murray Gell-Mann and Francis Low (1954) on  Quantum Electrodynamics (QED) and its development by Nikolai Bogolioubov and Dmitri Shirkov (1959), proceeded to extend their approach to strong interactions, for which at the time there was no generally accepted theory. 

Gell-Mann and Low's idea had been to perform the renormalization of electric charge by inserting an arbitrary momentum value $\lambda$ in place of the electron mass $m$. In this way, they were able to derive  a generalized expression for the renormalized charge $e_\lambda$, which  would be called today the running coupling constant. The parameter $e_\lambda$ obeyed the renormalization group equation for $m = 0$ as follows:

\begin{equation}  
\dfrac{d e_\lambda^2}{d (ln \lambda^2)} = \psi (0, e_\lambda^2) 
\end{equation}

\noindent where the function $\psi$ depended on the theory to be renormalized. 

Gell-Mann and Low concluded from their analysis for QED that the renormalization group equation did not provide any constraint on the measurable value of $e$. However, Wilson was convinced that ``startling consequences" could follow when applying the renormalization group equation to strong interactions (Wilson 1971, 1819). What he meant was the possibility to determine {\it a priori} the value of the strong coupling constant, and to show how this might happen was the main goal of his paper. The idea of deriving the values of coupling constants from purely theoretical considerations may seem strange today, but was still quite common at the time, as it had constituted the central tenet of Geoffrey Chew's bootstrap theory, which had dominated the theoretical stage in particle physics during the 1960s (Pickering 1984b, 73-78). According to that approach, all coupling constants and masses would generate themselves from a fundamental theory where all parameters were zero. Their values should in principle follow from conditions of analyticity and pole structure of the S-matrix, the function from which all scattering amplitudes are derived. Chew and his collaborators were never able to deliver a full theory, but their idea of deriving physical information from mathematical constraints like analiticity and pole structure were at the time very influential, and in his paper Wilson was following a similar, albeit much more limited program, by taking the renormalization group equation as starting point. The connection to the bootstrap is suggested by Wilson himself, who calls a key result of his analysis a ``bootstrap condition for the renormalized coupling constant" (Wilson 1971, 1831). At this point, let us summarize  the key arguments and results in Wilson's text which are relevant for the aim of this paper.

Wilson's approach was to qualitatively analyze the asymptotic behavior of the solutions to the renormalization group equation using methods from non-linear mechanics and electric circuit theory, introducing into high energy physics techniques of visualization which would soon become standard (Peskin 2014, 658).  At the beginning, he listed the different possible asymptotic behaviors of the solutions, classifying them into three types: (a) fixed points; (b) limit cycles and  (c) other types of behavior ``not easily characterized" (Wilson 1971, 1825). For simplicity, he chose to disregard the third kind of solutions, as they were ``more difficult to analyse" (Wilson 1971, 1825). Moreover, he only briefly discussed limit cycles, focusing almost exclusively on solutions displaying a fixed point for low and high momenta. Wilson was quite honest about the tentative character of his approach and the simplifying assumption he had to make in order to explore the mathematical features of the renormalization group equation and its solutions. 

Let us now follow Wilson's analysis. Starting from the assumption of a fixed-point asymptotic behavior, he further supposed that the function $\psi(0, x)$, which from now on we will for simplicity refer to as $\psi(x)$, had a very simple form: a function oscillating above and below zero, crossing the x-axis at four points:  $0, x_1, x_2$ and $x_3$. With the help of further simplifying assumptions, he argued that the solutions had the form shown in figure 1 (Figure 3 from Wilson 1971, 1827): monotonically increasing or decreasing functions which, for $\lambda$ going to zero or infinite, tended to a value equal to one of the four zero's of $\psi$. Wilson labeled these fixed points ``infrared" or ``ultraviolet" depending on whether they were limits for $\lambda$ going to zero or infinite. These functions were qualitative templates, providing the basis for further analysis. 
\begin{figure}[hbtp]
\caption{Figure 1: Qualitative behavior of the solutions for the zero-mass renormalization-group equation for $e_\lambda^2=$ plotted vs $ln\lambda^2$ (Figure 3 from Wilson 1971, 1827).}
\centering
\includegraphics[scale=1]{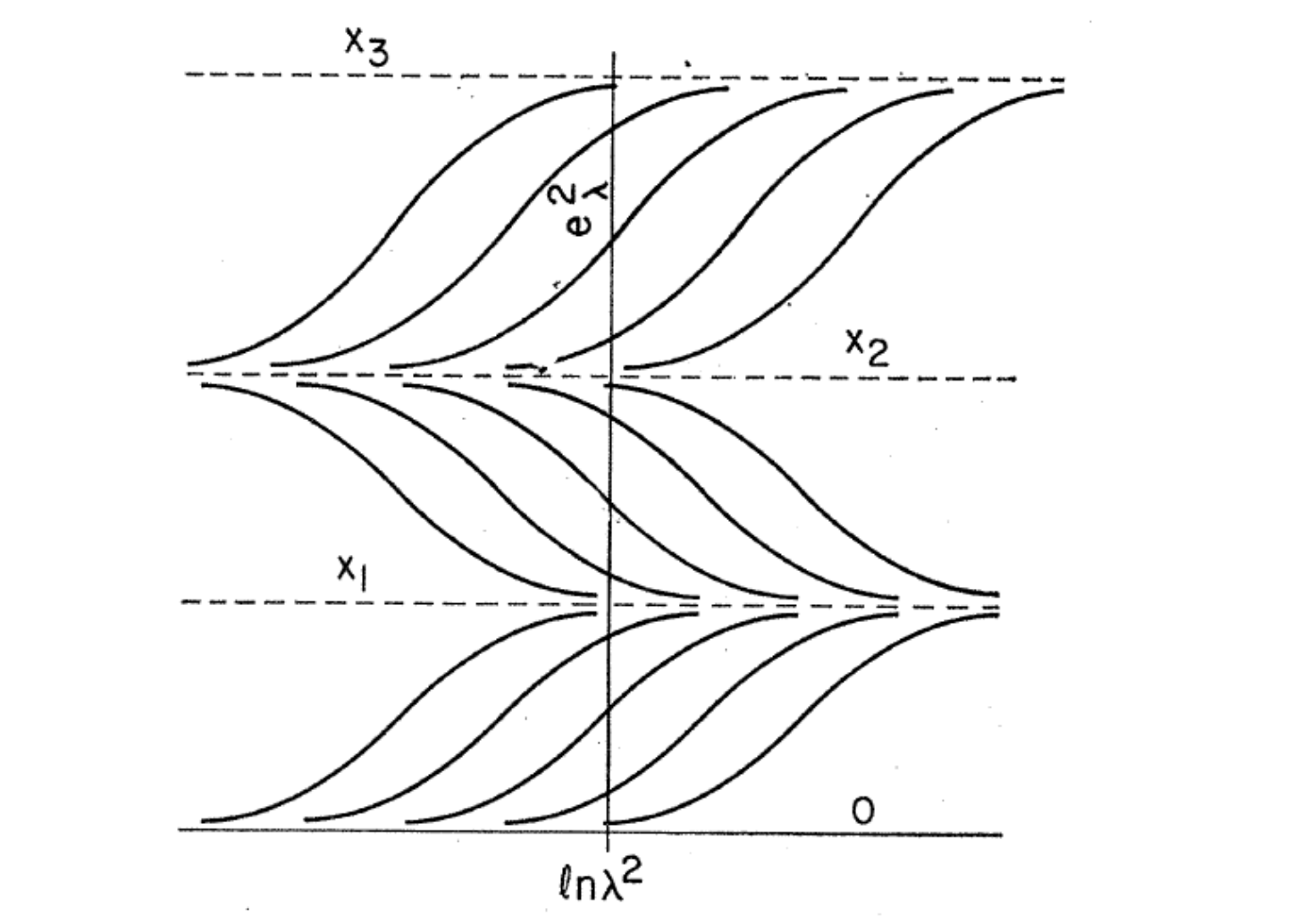}
\end{figure}

At this point, Wilson made a further distinction: either the strong interaction theory and its renormalization group equation were valid for all momentum values, or only for momenta $\lambda << \Lambda$ , where $\Lambda$ was a cutoff due to the appearance of other forces which could not be neglected, such as weak or electromagnetic interactions. The first case corresponded to the situation discussed by Gell-Mann and Low for QED, where the renormalization group had no predictive potential and one had to insert by hand the value of the physical coupling constant.  The second case was particularly interesting for Wilson, because he was convinced that, under those circumstances, the value of the strong coupling constant could be predicted \textit{a priori}. This was the ``startling'' result announced at the beginning of the paper. How was this possible? Here below, in a nutshell, Wilson's argument.

Wilson noted that, under the assumption that the renormalization-group equation was only valid for 
$\lambda << \Lambda$,  that equation would give no information whatsoever about the value of $e_\lambda$ when $\lambda$ approached the cutoff. Therefore, he proposed to regard the value of $e_\lambda$ at the cutoff as a random number lying between the two ultraviolet fixed points $x_1$ and $x_3$ (Wilson 1971, 1830). 
What happened then was the following: for $\lambda \rightarrow 0 $ the renormalization-group equation was valid and constrained  $e_\lambda$ to go towards the infrared fixed point $x_2$, as shown by figure 1. Since the value of $x_2$ was fully determined by the form of the function $\psi$, it was in principle possible to exactly predict the value of the strong coupling constant for $\lambda \sim 0$.\footnote{If the simplifying assumption that $m \sim 0$ was dropped, then the value was predicted for $\lambda \sim m$.} Wilson called this a ``bootstrap condition for the renormalized coupling constant" (Wilson 1971, 1831)

Having illustrated in a simple case the "startling" result, Wilson went on to further explore the implications of the bootstrap condition. He included masses in the renormalization group by defining dimensionless, generalized mass parameters equal to $\dfrac{m_\lambda}{\lambda }$ (Wilson 1971, 1839). At this point he introduced symmetry considerations, noting how the renormalization-group equation respected symmetry, because, if a symmetry-breaking parameter was zero for some value of $\lambda$, it would remain zero for all values. Because of this, it made sense to  distinguish between symmetry-breaking and non-symmetry-breaking parameters.

The symmetry-breaking parameters Wilson focused on were not masses, but rather coupling constants breaking the SU(3) quark symmetry.\footnote{It must be recalled that Wilson was writing well before the establishment of Quantum Chromodynamics as a theory of strong interactions, and that he was not making any specific assumption on the form of the strong Lagrangian. However, since it was known that strong forces had a SU(3) symmetry (the quark model) broken at low energy, the unknown Lagrangian must have an SU(3)-symmetry and contain smaller, SU(3)-breaking terms.}  These parameters were known to be of the order of 1 for momenta between 1 MeV and 1 GeV, but small for momenta much larger than 1 GeV. On this basis, Wilson assumed that the parameters varied monotonically as in Figure 1, and that they would therefore decrease when $\lambda$ increased, to become ``very small indeed" around the cutoff $\Lambda$ (Wilson 1971, 1839). In this regard, he remarked:

\begin{quote}

It is hard to see how this can come about unless these couplings also break an electrodynamic symmetry\footnote{Here Wilson assumed that electromagnetic interactions were those responsible for the fact that at the cutoff $\Lambda$ strong interactions could not be anymore studied in isolation, and that therefore their renormalization group equation was not valid anymore at those high momenta.}  [...] This means that there must be a symmetry common to electrodynamics and strong interactions which is broken by [these couplings] (Wilson 1971, 1839-40).

\end{quote}

\noindent Although, with hindsight,  this argument might be interpreted in terms of naturalness (in the sense connecting the notion with symmetry, as we will see), when situated in its context it appears to be just one more plausibility assumption on the behavior of the function studied, and not a claim that there was a physical reason why the coupling could not take on certain values. This was an opposite line of argument with respect to 't Hooft's \textit{a priori} request that only symmetry-breaking parameters be allowed to take very small numerical values. In Wilson's case, the pre-existing mathematical apparatus of the renormalization group equations suggested constrains on the numerical value of its solutions for given momentum range.\footnote{More precisely, parameters known to break at low momenta an internal symmetry of strong interactions were expected to also break a symmetry common to strong interactions and to the forces that became relevant at the cutoff $\Lambda$, for example electromagnetism. This conclusion did not follow necessarily from the analysis, but appeared plausible when considering that the same function of momentum was expected to satisfy two renormalization group equations: the one of strong interactions at lower energy, and the one for the theory valid above the cutoff at higher energies. The former equation required the parameters to become smaller and smaller when momenta increased, and the easiest way in which a seamless connection between the two ranges could take place, Wilson assumed, was that the two theories shared a symmetry broken by the parameters in question, which guaranteed that they would only receive small radiative corrections when nearing the cutoff.} In t'Hooft's scheme, an assumption for which no motivation was given (a ``dogma" ('t Hooft 1980, 136)) constrained the mathematical form of possible theories. 

However, so far Wilson had not yet discussed mass terms, but only symmetry-breaking terms in general, so we have to follow his reasoning further. For what regards the non-symmetry-breaking parameters, he used similar arguments to claim that it was ``likely" that they would receive large renormalization contributions at high $\lambda$, but smaller corrections for small $\lambda$. From this observation he concluded that the only parameters that could have large corrections for small $\lambda$, and thus break scale invariance, were those also breaking some symmetry of strong interactions theory. 
This applied also to generalized mass terms, since they always break scale invariance at small $\lambda$. In Wilson's words:
 
\begin{quote}
 
The breaking of scale invariance at low momenta is due entirely to couplings which also break internal symmetries; in particular, all generalized mass terms must break an internal symmetry (Wilson 1971, 1840).  
\end{quote}

\noindent But what about mass terms that were known not to be symmetry-breaking, such as scalar ones? Wilson's reasoning appeared to suggest that their coupling should be zero at all momenta, which amounted to a claim that no massive scalar particles would exist:

\begin{quote}
It is interesting to note that there are no weakly coupled scalar particles in nature; scalar particles are the only kind of free particles whose mass does not break either an internal or a gauge symmetry (Wilson 1971, 1840). 
\end{quote}

\noindent Thus, Wilson here pointed out that the physical prediction he had deduced from the mathematics of the renormalization group equation appeared empirically confirmed: no elementary scalar had so far been observed. At this point, combining these remarks with the previous analysis of symmetry-breaking terms, he came to the conclusion that mass terms also had to break a symmetry of the theory valid at the cutoff. In Wilson's words:

\begin{quote}

This discussion can be summarized by saying that mass or symmetry breaking terms must be 'protected' from large corrections at large momenta due to various interactions (electromagnetic, weak or strong). A symmetry-breaking term [...] is protected if, in the renormalization-group equation the right-hand side is proportional to [the coupling constant of the term] (Wilson 1971, 1840).

\end{quote}
 
\noindent Here we got back, finally, to the quote we started from. It took quite a long detour through Wilson's intricate, original, and today often problematic reasoning to arrive at it. However, reading the quote after having become more familiar with its context, it is difficult, if not impossible, to interpret it in terms of some principle of naturalness, as often done in the literature. On Wilson's part -- and this is the important point for the analysis presented here --  it was rather a reflection on how some solutions of the renormalization-group equation might behave under certain assumptions. 

The distinction between an (approximate) argument derived from an existing mathematical apparatus and a criterion based on an \textit{a priori}  assumption of physical or philosophical character is a very important one. In his analysis of Wilson's paper, for example, Porter Williams claims that Wilson in some way foreshadowed 't Hooft's or Susskind's naturalness, because he had allegedly recognized scalar particles as ``uniquely problematic on primarily physical grounds" (Williams 2018, 10). However, as shown above, Wilson did not argue that scalar masses were problematic on physical grounds, but rather on mathematical ones, i.e. that their behavior as a function of momentum would not fit the constraints dictated by renormalization group equations. 
This was the same kind of argument deployed by Wilson to show that, under given circumstances, one might predict the value of the strong coupling constant simply on the basis of the allowed form of functions solving the relevant renormalization group equations. As noted above, the idea that the mathematical structure of quantum field theory might by itself constrain measurable values of parameters had much in common with the notions at the core of bootstrap theory.
Like the bootstrap hypothesis, some of Wilson's reflections do not appear much plausible today, and he himself later rejected the validity of his reflections on scalar particles.\footnote{In 2005 Wilson wrote that this paper contained ``three bluders": the statements on scalar particles, the failure to recognize the possibility of asymptotic freedom and the idea that  limit cycles might be physically significant (Wilson 2005, 12). Interestingly Peskin in 2014 also said that Wilson spoke of ``three errors", but beside limit cycles and asymptotic freedom, he counted the idea that one might predict coupling constants instead of the critique to scalar particles (Peskin 2014, 658-659).}

To conclude, let us highlight what we think are the main difference between Wilson's claims from 1971 and later ideas of naturalness:

\begin{itemize}

\item[$\bullet$]Wilson addressed the issue of scalar mass terms as a side remark in a long and complex argument focusing on other topics. His main aim was showing the possibility of predicting the values of coupling constants.

\item[$\bullet$]Wilson did not regard the hypothesis of a cutoff as a necessarily true premise, or even as the most plausible one, but only as one of two equally probable alternatives.

\item[$\bullet$]Wilson did not apply any aesthetic or philosophical principles regarding nature, but rather a series of disparate, situated plausibility assumptions on how the solutions to specific equations might or not behave. He did this to extract physical meaning from the formalism, and not to impose physical principles onto mathematical models.

\item[$\bullet$]All results obtained were regarded as hypotheses to be empirically tested, and not as evidence in favor of a model. For example, since at the time there appeared to be no elementary scalar particles in nature, this fact could be interpreted as an empirical indication of the validity of the assumptions, and not the reverse!

\end{itemize}

\section{The rise of naturalness in the early 1980s}

As we saw in the previous Section, neither a naturalness principle nor a naturalness problem appeared in Wilson's 1971 paper. But, surely, the 1970s set the stage for the emergence of the naturalness problem around 1980. In that period, first the Weinberg-Salam model and then Quantum Chromodynamics became established as the theories respectively of electroweak and strong interactions. At the same time, the first speculative theories of physics at very high energies emerged, such as Grand Unified Theories. It was in discussions about the plausibility of these and other visions of new physics at high energies that different versions of an alleged naturalness problem of the Standard Model were proposed, for which one or the other  model of new physics had a solution.  The following papers, which started this trend, are today quite famous:

\begin{itemize}

\item[1)] Susskind's ``Dynamics of spontaneous symmetry breaking in the Weinberg-Salam theory" (1979);

\item[2)] 't Hooft's ``Naturalness, chiral symmetry, and spontaneous chiral symmetry breaking"(1980);

\item[3)] Veltman's ``The infrared-ultraviolet connection"(1981).

\end{itemize}

\noindent As said, these papers have been often discussed in the literature.\footnote{Discussions of one or more of these texts can be found for example in: Borrelli 20915, Dine 2015, Giudice 2008, Grinbaum 2012, Wells 2015.} Here we will just point out how  three  different, and not necessarily compatible definitions of naturalness were proposed, all with the aim of pinpointing some problem with the Standard Model and discuss its possible solutions. Schematically:

\begin{itemize}

\item[1)] Susskind spoke of ``a concept of naturalness which requires the observable properties of a theory to be stable against minute variations of the fundamental parameters" and  which was violated by the quadratic divergences in the renormalization of the Higgs mass (Susskind 1979, 2619). He never explained why this principle of naturalness should hold, but simply thanked Wilson for this insight  and went on to propose a ``natural" model for electroweak interactions based on a composite Higgs (Susskind 1979, 2625). Although Susskind thanked Wilson for eplaining him how scalar particles required ``unnatural adjustements of the bare constants", Wilson had made no such claim in 1971, as we saw (Susskind 1979, 2624).

\item[2)]'t Hooft explicitly characterized naturalness as a ``dogma" for which  no justification had to be offered ('t Hooft 1980, 136). His notion of naturalness was very different from Susskind's, and put symmetry center stage: ``at any energy scale $\mu$, a physical parameter or set of physical parameters $a_i (\mu) $ is allowed to be very small only if the replacements of $a_i (\mu) = 0 $ would increase the symmetry of the system" ('t Hooft 1980, 136). 't Hooft argued that the Higgs boson mass made the Weinberg-Salam model unnatural, but for different reasons than  those proposed by Susskind ('t Hooft 1980, 139-140). The bulk of 't Hooft's paper was devoted to the question of whether a model with composite Higgs as proposed by Susskind and his collaborators was natural in this new sense, and his conclusion was that it was not, but could perhaps be made natural if also fermions were regarded as composites.

\item[3)] Veltmann offered yet another definition of naturalness with the aim of finding and solving a problem arising within the Standard Model. He started from Susskind's definition of naturalness, and criticized it in that it was dependent on the specific regularization technique chosen. He noted how, in dimensional regularization, terms leading to quadratic divergences were set to zero. ``A naive person - Veltman remarked - could conclude that there are no quadratic divergences" (Veltman 1981, 447). Accordingly, Veltman reformulated the naturalness criterion to make it independent of the regularization method as follows: ``radiative corrections are supposed to be of the same order (or much smaller) than the actually observed values" (Veltman 1981, 446). In Veltman's scheme, as in 't Hooft's one, symmetry played an important role, because it kept radiative corrections small. He concluded the paper with the suggestion that perhaps supersymmetry might offer a solution to this newly defined naturalness problem of the Standard Model (Veltman 1981, 451). 

\end{itemize}

\noindent As this brief overview of the seminal naturalness papers shows, around 1980 the naturalness principle was a vague, ambiguous notion for which different, largely incompatible definitions existed.  For example, while for Veltmann radiative corrections were the starting point to define naturalness, they played no role in 't Hooft's notion. Symmetry was essential for t'Hooft's naturalness, but found no mention in Susskind's.  Yet this ambiguity did not appear as a problem, and no discussion on which of these (and various later) versions of naturalness was the correct one took place. The important feature shared by all definitions was that they pointed to some problem of the Standard Model and offered a tentative solution for it. As noted in Section 1, this flexibility would soon prove to be an asset rather than a flaw because, by choosing different definitions of the naturalness problem, theorists could point to different answers and so promote one or the other specific model of new physics. The best-know case of this kind was the connection to supersymmetry of Susskind's specific definition of naturalness in terms of quadratic divergencies.

As seen from today, the connection between naturalness and supersymmetry has a long tradition. Yet,  although this connection motivated the development of supersymmetry,  that approach did not emerge as an answer to the naturalness problem, as supersymmetric models of particle interactions were already on the market since the 1970s. At the same time, as we saw above, the origin of naturalness was not linked to supersymmetry, but rather to composite Higgs models. However, soon after the idea of a naturalness problem of the Standard Model linked to quadratic divergences became known, people working on supersymmetry started pointing out that those divergences were absent in supersymmetric theories. Veltman appears to have been the first to do so in print, but he was soon followed by others.

A significant example of how supersymmetry and naturalness were virtuously combined is the work by John Ellis, Mary Gaillard and Bruno Zumino. In 1980 the three physicists proposed a supersymmetric GUT model unifying particle interactions with gravity and assuming the existence of elementary preons (Ellis, Gaillard and Zumino 1980). However, in that case, they did not employ  any naturalness argument. One year later, though, they wrote a longer review on ``Superunification" in which they offered what they called a ``recital of the problems which supersymmetry may be called upon to solve" and included among them the quadratic divergences of the Higgs mass (Ellis, Gaillard and Zumino 1981, 2-3, 9-10). The final aim of the paper was to propose essentially their 1980 model as a solution to those various problems, among which the naturalness one. Thus, defining naturalness in terms of the absence of quadratic divergences showed that supersymmetry was in an ideal position for making the Standard Model natural. 

From the early 1980s onward the synergy between supersymmetry and naturalness grew, yet naturalness remained a vague, flexible resource which theorists working along different research  lines could tap to motivate and guide their own work. Naturalness had stepped out from the specialist technical niche were it was born, to become one of the many overarching ideals of physical research, such as unification or simplicity.

\section{Intermezzo: Naturalness becomes popular}

As far as we know, the paper ``Naturalness in Theoretical Physics'' by Philip Nelson, which appeared in 1985 in the magazine {\it American Scientist}, is one of the first popular discussions of the meaning and role of naturalness. This is confirmed by the author himself:  at the time he is writing, he claims, ``naturalness seems to be one of the best kept secrets of physicists from the public, a secret weapon for evaluating and motivating theories of the world on its deepest levels" (Nelson 1985, 60). His article was therefore aimed to share this ``secret'' with the large public. 

Given the context, the paper's scope was quite broad. The incipit of the paper makes it explicit, by placing the concept in the frame of what is called, in philosophical jargon, a situation of underdetermination: i.e., the case where there are more than one theory justifying or ``saving''  the same phenomena. Starting with noting that ``theoretical physics is not what it used to be", because ``less determined by experiment as before'', Nelson's question was how to ``distinguish good theories from bad ones'' (Nelson 1985, 60). In such cases, philosophers typically say that theory choice is oriented by the so-called theoretical or extra-empirical virtues, such as simplicity, 
beauty, unification, and so on. On his part, Nelson situated naturalness among the principles helping us in formulating good theories. Other such principles were, for Nelson, the cosmological principle (``our position in the cosmos is completely undistinguished''), the principle of insulation (``succeeding scales are insulated from one another'', Nelson 1985, 62), and the various symmetry principles (Nelson 1985, 61-4).

What is the status of all these principles? For Nelson, they ``were arrived arrived at by dint of hard work and much trial and error'' and ``the fact that they have become dogma today rests not so much on their intrinsic ``beauty'' as on their pragmatic successes'' (Nelson 1985, 60). Successes which should not be taken as an indisputable guarantee for the future, Nelson was careful to add: naturalness can also ``give poor counsel'' sometimes (Nelson 1985, 60).

In fact, the issue of the status of this sort of principles is very general and much discussed in the literature, especially among historians and philosophers of physics (not necessarily putting all such principles in the same category).\footnote{See for example (Williams 2015) and references therein.} Leaving this general discussion aside, let us just have a closer look at Nelson's account of naturalness (and why he thought it is worth making it popular). Indeed, in his account of ``numerical naturalness'' we can find the main features of what has become, today, the controversial ``principle of naturalness''.\footnote{Note that Nelson distinguishes between a ``structural naturalness'', which has essentially to do with simplicity (p. 60), and a ``numerical naturalness'', which is, in fact, the today's meaning of the concept. We will focus, therefore, only on his numerical naturalness.}

As said, Nelson's starting point was the underdetermination issue, with respect to which a ``strong naturalness problem'' is identified as follows (Nelson 1985, 61):

\begin{quote}

We have a strong naturalness problem whenever the set of theories which even remotely resemble our world is a tiny subset of all the acceptable theories.

\end{quote}

\noindent  The cure he proposed for this problem was ``by slicing the latter class down to size.. finding some new principles which render most of its members unacceptable'' (Nelson 1985, 61).  In this way, Nelson noted, ``theorists often permit the introduction of new structures into their theories'', typically new symmetries, acting as powerful constraints. The general idea was that ``given some mysterious special feature of the observed world'' (Nelson 1985, 62), like a mass zero (as in the case of the photon) or approximately zero (as for the pion),  there should be a symmetry which explains it (exact gauge symmetry for the photon, approximate chiral symmetry for the pion). Symmetry was thus ``the principle for reducing problems of numerical naturalness to questions of structure'' (Nelson 1985, 63). 
Nelson illustrated this function of symmetry by means of a number of case studies from particle physics and cosmology, where all the theoretical ingredients of the developments in fundamental physics up to the middle 80s were called into play (renormalization theory and effective field theories, scale hierarchy and grand unification, the problem of the Higgs mass, fine-tuning, ..). 

When discussing the various examples, Nelson repeatedly stressed how a ``naturalness problem'' or even a ``strong naturalness problem'' was present, and suggested new theoretical developments as solutions. For example, he stated that ``the Standard Model has a glaring naturalness problem'' (Nelson 1985, 64) that might be solved by unified theories. The asymmetry between matter and antimatter, too, posed a ``strong naturalness problem'' (Nelson 1985, 65) which unified theories might solve. However, he also noted that ``unified theories themselves suffer from new naturalness problems even as they solve old ones'' (Nelson 1985, 65). Here the function of naturalness in theoretical physics was not simply to guide the choice among theories in cases of underdetermination, but also to point at problems and prompt the search for new theories to solve them.

Towards the end of his paper, a whole Section was devoted to discussing the naturalness problem in relation to the Higgs mass. Nelson's approach was strongly inspired to 't Hooft's 1980 treatment, where symmetry plays a central role in characterizing naturalness, as recalled in Section 5.\footnote{As acknowledged by Nelson himself, at p. 66.} Interestingly, though, the Section has the title ``Wilson's criterion'', and in Nelson's description of Wilson's 1971 contribution, we even read: 
``Wilson's criterion states that a small parameter in an effective theory is acceptable only if setting it to zero yields a more symmetrical theory'' (Nelson 1985, 66). 
We can then well say that, at the time naturalness became popular, the misreading of Wilson's contribution was already effective.

\section{The era of naturalness... and beyond}

Although written for a non-expert audience, Nelson's paper offered a good picture of the role of naturalness in contemporary particle research. Naturalness was a flexible criterium that could be employed to highlight problems of the Standard Model and motivate the search for solutions. By the late 1980s there was a broad consensus that the Standard Model had a naturalness problem, although different characterizations of the problem (and its possible solutions) coexisted.

The relationship of naturalness and supersymmetry offers a good example of how naturalness could be appropriated and reshaped to serve the goals of the moment. As we saw above, this relationship had not been there in the beginning, but was soon established. People working on supersymmetry underscored how unnatural the quadratic divergences of the Higgs mass were, and how supersymmetry could let then disappear. Moreover, in 1988 Riccardo Barbieri and  Giudice took a further, very innovative step: in their paper on ``Upper bound on supersymmetry mass particles" they deployed naturalness to derive experimentally testable prediction (Barbieri and Giudice 1988).

In some more detail, Barbieri and Giudice provided a quantitative estimate of how much supersymmetric particles could be made heavier than their partners before the necessary fine-tuning became unnatural, by requiring that ``no cancellation takes place among the physical parameter of the minimal supergravity model by more than one order of magnitude" (Barbieri and Giudice 1988, 73). On this basis, they predicted that the lightest supersymmetric particle had a mass around 100-200 GeV, and so might be detected at the LEP collider which was soon due to start operating at CERN. 

In 1989, the LEP accelerator started working and in 1995 its first running phase (LEP-1) was concluded without any evidence of supersymmetric particles. In the same year, the paper ``Measures of fine-tuning" by George Anderson and Diego Casta\~no (1995) appeared, criticizing the estimate by Barbieri and Giudice as too restrictive and providing an alternative proposal.  They stated that ``the traditional prescription does not distinguish between instances of global sensitivity and real instances of fine tuning" (Anderson and Casta\~no 1995, 301), while their proposal took into account the specific features of the system studied:

\begin{quote}
[Barbieri and Giudice's] prescription is an operational implementation of Susskind's statement of Wilson's sense of naturalness, ``Observable properties of a system should be stable against minute variations of the fundamental parameters" [...] Our measure is an operational implementation of a modified version of Wilson's naturalness criterion: Observable properties of a system should not be unusually unstable against minute variations of the fundamental parameters (Anderson and Casta\~no 1995, 307).
\end{quote}

\noindent Changing from ``stable" to ``not unusually unstable" they could argue that the upper limits of supersymmetric particles where higher than assumed so far, and thus compatible with LEP-1 results (Anderson and Casta\~no 1995, 307).

In the following decades, naturalness continued to provide a central, if vaguely defined tool for model-building. For example, as shown by (Grinbaum 2012), between the 1990s and the beginning of the new millennium increasingly complex measures of fine-tuning were developed, while at the same time naturalness could be associated with the need for explaining the hierachy between the electroweak scale and the Planck mass. 
We need not go into the great variety of models of new physics which emerged in that period: naturalness continued to provide motivation and guidance both in theory and experiment.\footnote{For an overview and examples of approaches to model-building from the 1990s onward see (Borrelli 2012).} As we saw above, in 2008 it was regarded as a main motivation for the LHC and Giudice devoted to it an article in the volume on \textit{Perspectives on LHC Physics}.

However, by 2013 the first phase of LHC had confirmed all predictions of the Standard Model and failed to provide evidence for physics beyond it. It was at that point that a growing number of physicists started asking whether there was a naturalness problem at all, attempting to redefine or discard the criterion of naturalness: the "post-naturalness era" had begun.\footnote{Contributions to that debate include: Dine 2015, Giudice 2013 and 2017, Feng 2013, Hossenfelder 2018, Wells 2015, Williams 2015.} 

\section{Conclusion: Practices, problems, principles}

In which contexts was the term ``naturalness" employed from the 1970s until today? Which practices were referred to as natural or unnatural? Which goals were the theorists pursuing when using naturalness arguments? These are the issues we have tried to address in this paper.

In particular, although attempts have been made to see some steps in the development of the Standard Model as successes of naturalness, we aim to show with our analysis that, in the 1970s, the use of ``natural" or ``naturalness" cannot be interpreted as what would be later indicated by those terms. This is also, and especially, the case with the famous paper by Wilson (1971) which has often been indicated as marking the origin of naturalness. In that period, people characterized plausibility assumptions as ``natural" without necessarily seeing them as expressing some principle of nature.  As we saw, Wilson explored the formal properties of the renormalization group equation and its solutions using methods taken over from classical linear mechanics and the study of electric circuits. By making assumptions on the behavior of fixed-point solutions, he strove to derive consequences with physical relevance, most notably an \textit{a priori} determination of the value of the strong coupling constant.  Methodologically, his approach displayed similarities to bootstrap theory. In this context Wilson also argued that, under specific conditions (fixed-point solution, existence of a cutoff), scalar mass terms would have a vanishing coupling. This claim has \textit{a posteriori} been seen as an early formulation or at least a foreshadowing of the later naturalness notion, but we have argued that this was not the case. Wilson was using a formal analysis of the mathematical apparatus of quantum field theory to derive physical consequences, and not imposing from the outside naturalness conditions on numerical values, as later authors would do.
Moreover, his assumptions were regarded as temporary, short-lived guidelines which would be soon proved or disproved by experiments, and not as strong arguments for or against some model.  

From the early 1980s onward, instead, practices emerged which can be interpreted as uses of naturalness in today's sense.   While Wilson had tried to derive physical predictions from mathematical reflections, authors working in the 1980s strove to determine the possible mathematical forms of theories by imposing conditions on their numerical predictions in the name of a physical principle, or a ``dogma`` of naturalness, for example by rejecting as unnatural very small numbers.  Naturalness was employed for non-empirical arguments of various character, making one model less plausible, and motivating some alternative even in absence of empirical tests. Things characterized as unnatural could be of various type: quadratic divergencies, fine-tuning, renormalization-group instability, or simply the ratio of two very different masses, like those of light and heavy quarks.\footnote{For a discussion of different versions of naturalness after the 1980s see (Borrelli 2015, Giudice 2008, 2013, 2017, Grinbaum 2012, Hossenfelder 2018, Wells 2015).}  

In this context, the goal in speaking of naturalness was twofold: on the one side, to argue that the Standard  Model had a naturalness problem; on the other side, to show that some specific model (technicolor, supersymmetry, extra dimensions, ...) could solve it. There was no universally accepted, clearly formulated naturalness principle guiding research, but only the shared conviction that a naturalness problem with the Standard Model existed and could manifest itself in several different ways. Naturalness was thus a flexible, heuristically productive tool for motivating and inspiring exploratory modeling in high-energy physics. In a sense, it was not the criterium of naturalness which shaped the goals of research, but rather the reverse. After the establishment of the Standard Model in the 1970s, theorists were confronted with the problem of choosing among a wide range of models of new physics, while experiments delivered no indications how it might look like. In this context, the formulation of different versions of the naturalness problem became a means to find guidance in the model-building enterprise. 

As long as the belief in a naturalness problem of the Standard Model held, the vagueness of naturalness was its main asset. Once the LHC results cast doubts on the naturalness problem, however, critical eyes turned to naturalness, and its asset now appeared as a liability, or at least as a feature in need to be reformed. Yet, looking back at the history of the practices of naturalness has provided a deeper understanding of its productive function in high energy physics research of the last decades, suggesting that the principle of naturalness was never so mighty as assumed afterwords. As situating this development has allowed us to show, naturalness was a conceptual tool for producing problems to solve, and not a principle to uphold at all costs.

\section*{Acknowledgements}
We are grateful to the participants in the 2018 Aachen Workshop on ``Naturalness, Hierarchy, and Fine Tuning" and to an anonymous referee for very helpful comments and suggestions. Arianna Borrelli wishes to acknowledge funding by the project ``Exploring the ``dark ages" of particle physics: isospin, strangeness and the construction of physical-mathematical concepts in the pre-Standard-Model era (ca. 1950-1965)" (German Research Council (DFG) grant BO 4062/2-1), and the Institute for Advances Studies on Media Cultures of Computer Simulation (MECS), Leuphana Universität Lüneburg (DFG research grant KFOR 1927).



\end{document}